\documentclass[aps,prb,reprint,superscriptaddress,
showpacs]{revtex4-1}
\usepackage{graphicx}
\usepackage{dcolumn}

\begin{document}

\title{Topological Phase Transition Under Pressure in the Topological Nodal Line Superconductor PbTaSe$_2$}

\author{C. Q. Xu}
\affiliation{Advanced Functional Materials Lab and Department of Physics, Changshu Institute of Technology,
Changshu 215500, China}
\affiliation{Department of Physics, Hangzhou Normal University, Hangzhou 310036, China}
\author{R. Sankar}
\affiliation{Institute of Physics, Academia Sinica, Nankang, Taipei R.O.C. Taiwan 11529}
\affiliation{Center for Condensed Matter Sciences, National Taiwan University, Taipei 10617, Taiwan}
\author{W. Zhou}
\affiliation{Advanced Functional Materials Lab and Department of Physics, Changshu Institute of Technology,
Changshu 215500, China}
\author{Bin Li}
\email[]{libin@njupt.edu.cn}
\affiliation{Information Physics Research Center, Nanjing University of Posts and Telecommunications, Nanjing, 210023, China}
\author{Z. D. Han}
\affiliation{Advanced Functional Materials Lab and Department of Physics, Changshu Institute of Technology, Changshu 215500, China}
\author{B. Qian}
\email[]{njqb@cslg.edu.cn}
\affiliation{Advanced Functional Materials Lab and Department of Physics, Changshu Institute of Technology,
Changshu 215500, China}
\author{J. H. Dai}
\affiliation{Department of Physics, Hangzhou Normal University, Hangzhou 310036, China}
\author{Hengbo Cui}
\affiliation{Condensed Molecular Materials Laboratory, RIKEN, Wako-shi, Saitama 351-0198, Japan}
\author{A. F. Bangura}
\affiliation{Max-Planck-Institut f\"{u}r Festk\"{o}rperforschung, Heisenbergstr. 1, D-70569 Stuttgart, Germany}
\author{F. C. Chou}
\affiliation{Center for Condensed Matter Sciences, National Taiwan University, Taipei 10617, Taiwan}
\author{Xiaofeng Xu}
\email[]{xiaofeng.xu@cslg.edu.cn}
\affiliation{Advanced Functional Materials Lab and Department of Physics, Changshu Institute of Technology,
Changshu 215500, China}
\affiliation{Department of Physics, Hangzhou Normal University, Hangzhou 310036, China}

\date{\today}

\begin{abstract}
A first-order-like resistivity hysteresis is induced by a subtle structural transition under hydrostatic pressure in the topological nodal-line superconductor PbTaSe$_2$. This structure transition is quickly suppressed to zero at pressure $\sim$0.25 GPa. As a result, superconductivity shows a marked suppression, accompanied with fundamental changes in the magnetoresistance and Hall resistivity, suggesting a Lifshitz transition around $\sim$0.25 GPa. The first principles calculations show that the spin-orbit interactions partially gap out the Dirac nodal line around $K$ point in the Brillouin zone upon applying a small pressure, whilst the Dirac states around $H$ point are completely destroyed. The calculations further reveal a second structural phase transition under a pressure as high as $\sim$30 GPa, through which a transition from a topologically nontrivial phase to a trivial phase is uncovered, with a superconducting dome emerging under this high-pressure phase.
\end{abstract}



\maketitle

\section{Introduction}
The recent development in topological physics has significantly extended the interest from an isolated Dirac node, either in 2D surface states or in 3D semimetals, to the lines or loops of Dirac nodes that are protected by the interplay of symmetry and topology\cite{XiaoHu,Kim,Mullen,a}. The Dirac nodal-line (DNL) semimetals have extended band touching of conduction and valence bands along a one-dimensional line in the momentum space and are expected to host a variety of exotic transport phenomena. Particular interest arises if these DNL semimetals also host superconductivity which are often regarded as strong candidates of topological superconductors (TSCs) whose low-lying excitations may be described by Majorana Fermions, defined as fermions which are their own antiparticles that are proposed to exist at the edge of a TSC\cite{Hasan10}. However, the material realization of the TSCs is especially rare, in particular for ones with stoichiometric compositions. For example, although the topologically protected surface states have recently been claimed in the centrosymmetric superconductor $\beta$-PdBi$_2$ by angular-resolved photoemission spectroscopy (ARPES), unambiguous evidence in favor of such topological states from other experimental techniques is still lacking\cite{Sakano15,Herrera,Kacmarcik,LuXin16}.

Among these TSC candidates, PbTaSe$_2$ is special, if not unique, in that its DNL states have been firmly identified by ARPES\cite{ncomms10556} and found to be rather robust against spin-orbit coupling (SOC), which often opens up a gap and induces topologically nontrivial band-inverted states\cite{Cava,sciadv}. In addition, its superconducting state exhibits many interesting properties, including the strong upward curvature in its $H_{c2}$ and a V-shaped pressure dependence of $T_c$ in the polycrystalline samples\cite{JianruiWang}. Further, its superconducting gap has been reported to be nodeless\cite{Jia,SYLi,Yuan}, meeting the
requirement of a topological superconductor. Recent pressure measurements on single crystals, however, reveal a pronounced resistivity hysteresis associated with a subtle structure modification and a drastic suppression of $T_c$ once $P$$\geq$0.25 GPa\cite{Canfield}. The questions remain on how the DNL features are modified by this structural transition and to what extent the topological states can survive when subjected to higher pressures.

In this work, we report experimental transport measurements up to 2 GPa on high-purity single crystals of PbTaSe$_2$, combined with the first principles calculations on its electronic structure up to 60 GPa, with a special emphasis on its topological features. Our experiments reveal a marked resistivity hysteresis upon the application of a small pressure. The superconducting $T_c$ is quickly suppressed above $\sim$0.25 GPa, accompanied by a sudden change in other transport coefficients, like residual resistivity, magnetoresistance and Hall resistivity. The calculations suggest this low-$P$ phase transition at $\sim$0.25 GPa is associated with a subtle change in its structure that opens up a small energy gap in a part of the DNL structure in the presence of SOC. Our calculations further uncover a phase transition from topological nontrival bands to topological trivial states, induced by a second structure transition under $\sim$30 GPa. Under this high-$P$ phase, a superconducting dome is resolved on the assumption of pairing from electron-phonon interactions, which motivates future high pressure measurements in due course.

\section{Experiment}
High-quality single crystals of PbTaSe$_2$ were grown by chemical vapor transport method. The detailed process of synthesis was described in Ref. [\cite{Sankar,Sankar17}]. Magneto-transport measurements were performed by a standard four-probe lock-in technique in a Quantum Design PPMS equipped with a 9 Tesla magnet. For the hydrostatic pressure measurements, samples were loaded into a piston type pressure cell and the actual pressure was determined by measuring the superconducting transition temperatures of Pb. Daphne 7373 oil was applied as the pressure transmission media. For the data under different pressures, the same contacts were used throughout the measurements such that the geometric errors in the contact size were identical for different runs.

To determine high-pressure structures, we used the evolutionary crystal structure prediction method USPEX~\cite{ux1,ux2,ux3}.
Predictions were made from ambient pressure up to 60 GPa with a step of 5 GPa. Structure relaxations, enthalpy and electron-phonon calculations were performed using the Quantum-ESPRESSO~\cite{QE} code with ultrasoft pseudopotential method and the plane wave basis. The cutoffs were chosen as 30 Ry for the wave functions and 300 Ry for the charge density. The generalized-gradient approximation of Perdew-Burke-Ernzerhof (PBE)~\cite{GGA} was used for the exchange-correlation energy function. The electronic structure calculations with high accuracy for the stable structures were performed using the full-potential linearized augmented plane wave (FP$-$LAPW) method implemented in the WIEN2K code.~\cite{Wien2k} The generalized gradient approximation (GGA)~\cite{GGA} was applied to the exchange-correlation potential calculation. The muffin tin radii were chosen to be 2.5 a.u.\ for Pb and Ta, and 2.37 a.u.\ for Se. The plane-wave cutoff was defined by $RK_{max}=7.0$, where $R$ is the minimum LAPW sphere radius and $K_{max}$ is the plane-wave vector cutoff. Spin-orbit coupling was included in the calculations.

\section{Results}

\begin{figure}
\includegraphics[width=8.2cm]{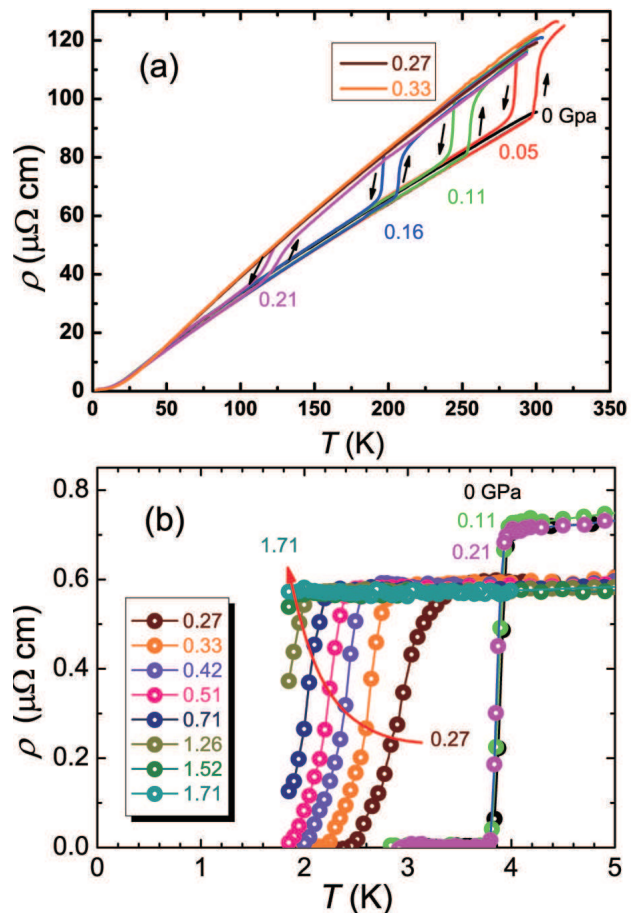}
\caption{\label{RT} (a) Temperature dependence of resistivity under different pressures. (b) An enlarged view of the resistivity curves at low temperatures.}
\end{figure}

Figure \ref{RT}(a) illustrates $\rho(T)$ profiles under various hydrostatic pressures ($P$). At ambient pressure, the $T$-dependence of resistivity is typical of those reported in the literature\cite{Jia,SYLi,Yuan}. Upon the application of a tiny pressure, e.g. 0.05 GPa, a sudden drop occurs around 286.5 K on the $\rho(T)$ curve during the cooling process. When warming up, the corresponding resistivity anomaly slightly shifts to a higher temperature, resulting in a pronounced hysteresis. Similar resistivity hysteresis has recently been reported by Kaluarachchi \textit{et al.}, which was ascribed to a subtle structural modification from the high-$T$ $P\bar{6}m2(1e)$ phase to the low-$T$ $P\bar{6}m2$ phase (see Fig. [\cite{Canfield}] for the corresponding structures). At ambient pressure, for example, with increasing $T$, the room temperature $P\bar{6}m2$ phase changes to $P\bar{6}m2(1e)$ which can be obtained by shifting the Pb atom from the 1a-Wyckoff coordinate in $P\bar{6}m2$ to the 1e-Wyckoff position without changing its overall symmetry. According to Kaluarachchi \textit{et al.}, this structure transition takes place at 425 K under ambient pressure, i.e., a temperature higher than our measurements\cite{Canfield}. Compared with $P\bar{6}m2$, the high-$T$ $P\bar{6}m2(1e)$ phase displays an obvious contraction in the $c$-axis length while the $a$-axis is slightly expanded\cite{Canfield}. Here, we define the structural transition temperature $T_s$ as the average of the characteristic temperatures of the resistivity anomalies on the cooling and heating processes. With increasing pressure, $T_s$ is found to be fast suppressed and disappears at $P_c$$\sim$0.25 GPa. Remarkably, the pressure suppression rate for $T_s$ reaches as large as $\sim$440 K/GPa, signifying a highly sensitive pressure response.

\begin{figure}
\includegraphics[width=8.2cm]{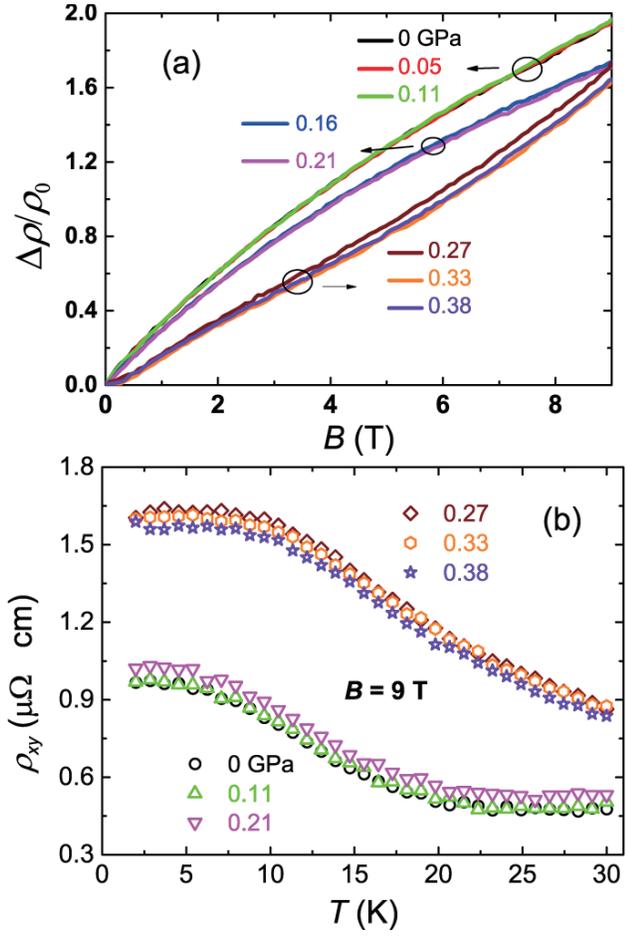}
\caption{\label{MRhall} (a) Magnetic field dependence of the magnetoresistance ($\triangle\rho/\rho_0$) under different pressures. (b) Temperature dependence of Hall resistivity $\rho_{xy}$ for different pressures.}
\end{figure}

In Figure \ref{RT}(b), a blow-up of resistivity in the low-$T$ regime clearly split into two branches. At the low pressure region with $P \leq 0.21$ GPa, the sample shows sharp superconductivity with $T_c$ around 3.9 K and $T_c$ varies little with pressure. A notable suppression of $T_c$ is clearly seen when $P$ increases above 0.27 GPa. This suggests two superconducting phases are separated around $P_c$$\sim$0.25 GPa. It is worth noting, this critical pressure is the same as the pressure where the resistivity hysteresis disappears, implying that these two superconducting phases correspond to different crystal structures. Besides, the residual resistivity, defined as the resistivity at 5 K, also shows a sudden drop across this critical pressure.

To gain further insight into the underlying electronic changes across this critical pressure, the magnetoresistance and the Hall resistivity are studied, as shown in Figure \ref{MRhall} (a) and (b), respectively. The latter has not been reported thus far. At low pressures, the field dependent MR shows an interesting downward curvature, analogous to many recently-discovered topological semimetals. With increasing $P$, however, the size of MR is slightly reduced and the shape of MR becomes superlinear in field. Similar drastic changes across the critical pressure are also observable in the $T$-dependent Hall resistivity ($\rho_{xy}$) curves. When $P$$\geq$$P_c$, the Hall resistivity is suddenly elevated, consistent with a pronounced change in the underlying electronic properties induced by the structural transition.

\begin{figure}
\includegraphics[width=8.2cm]{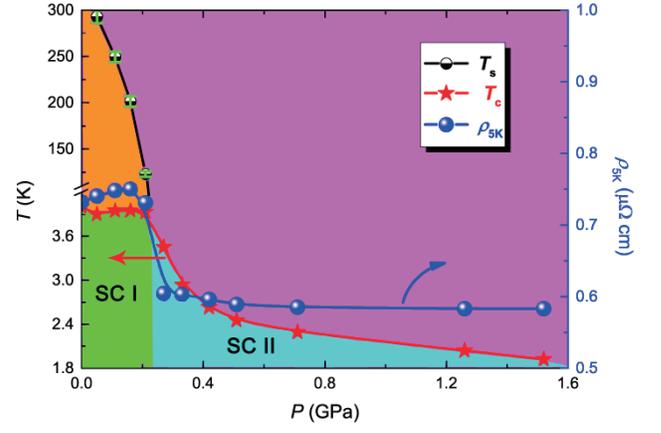}
\caption{\label{phasediagram} Phase diagram for the pressure dependence of the structural transition temperature $T_s$, the superconducting transition temperature $T_c$, and the resistivity at 5 K ($\rho_{\text{5K}}$).}
\end{figure}

Based on the above transport data, the pressure dependent phase diagram is summarized in Figure \ref{phasediagram}. In this phase diagram, two distinct superconducting phases separated by a structural transition are resolved. The quick residual resistivity drop, along with the abrupt changes in the magnetoresistance and the Hall effects, clearly indicates the remarkable change in the electronic properties across the phase boundary.

\begin{figure}
\includegraphics[width=8.2cm]{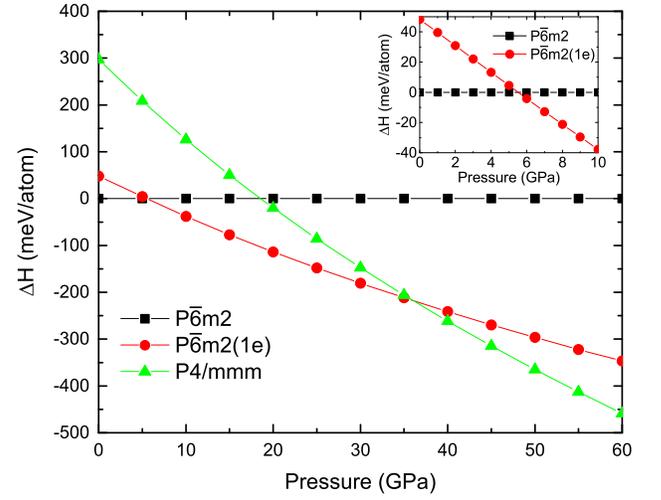}
\caption{\label{enthalpy} Calculated enthalpies per atom as a function of pressure from 0 to 60 GPa with respect to
$P\bar{6}m2$ structure. Inset shows the enlarged enthalpy comparison between $P\bar{6}m2$ and $P\bar{6}m2(1e)$ from 0 to 10 GPa.}
\end{figure}

\begin{figure*}
\includegraphics[width=18cm]{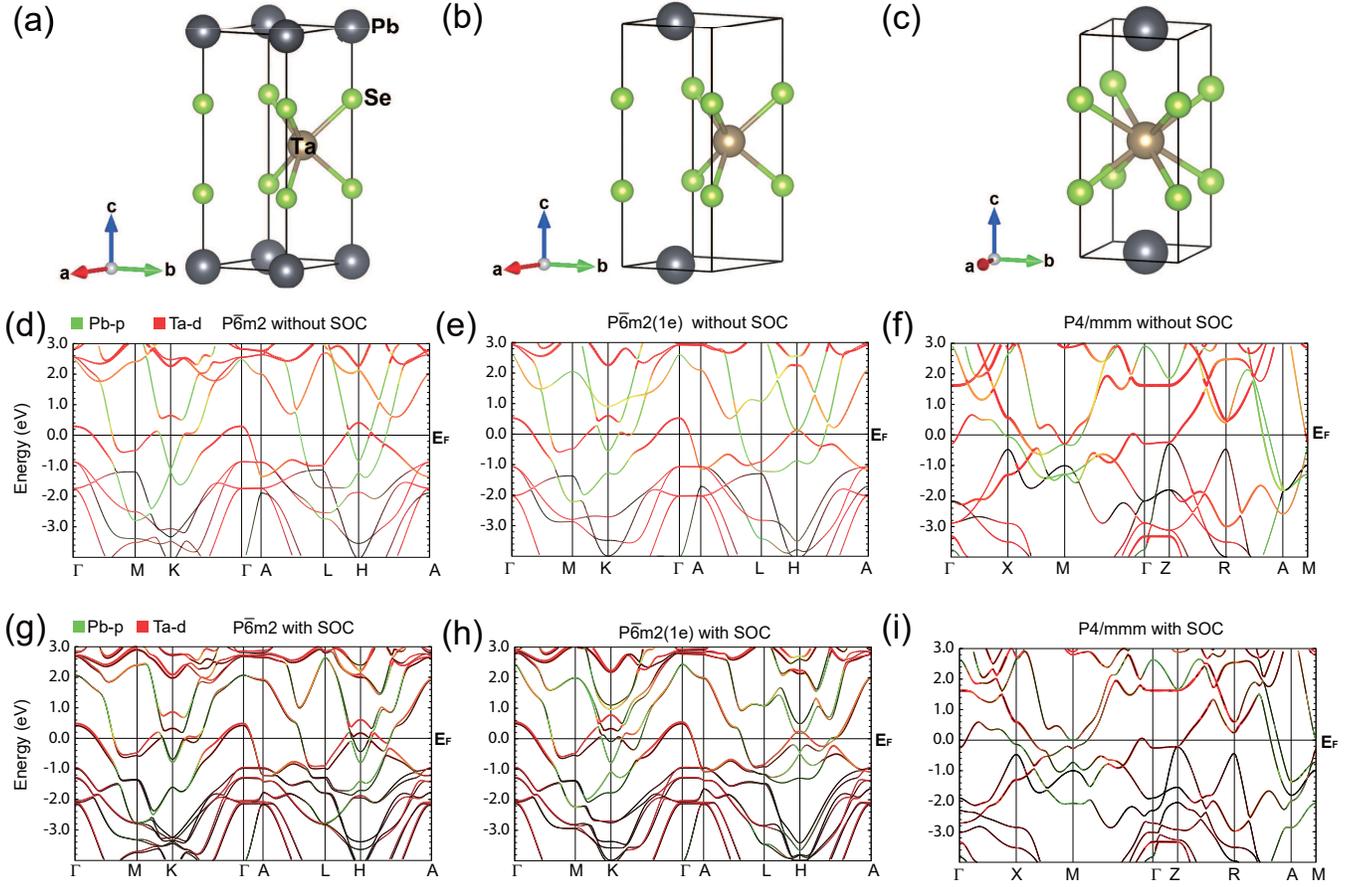}
\caption{\label{bands} The schematic unit cell for (a) non-centrosymmetric hexagonal $P\bar{6}m2$, (b) $P\bar{6}m2(1e)$, and (c) tetragonal $P4/mmm$ phases. The band structures of PbTaSe$_{2}$ calculated without (d)-(f) and with (g)-(i) SOC for three different structures.  The weights of the Pb $p$ and Ta $d$ orbital contribution are color-coded by green and red, respectively.}
\end{figure*}

To clarify the nature of transition between these two superconducting phases, detailed enthalpy calculations for different crystal structures are performed. The resultant enthalpy-pressure ($\Delta H$-$P$) curves are plotted in Figure \ref{enthalpy}, where the relative enthalpy differences $\Delta H$ are calculated with respect to the $P\bar{6}m2$ structure. We find that with increasing pressure, PbTaSe$_2$ undergoes two structural phase transitions, from the $P\bar{6}m2$ to $P\bar{6}m2(1e)$ structure at $\sim$5 GPa, then to the tetragonal $P4/mmm$ structure at $\sim$30 GPa. Below 5 GPa, the most stable structure is the $P\bar{6}m2$ which is the original phase found in experiment. As the pressure increases above $\sim$5 GPa, the $P\bar{6}m2(1e)$ structure becomes more stable. Note that the $P\bar{6}m2(1e)$ structure shares the same global symmetry with $P\bar{6}m2$, with only the Pb atom shifting from the $1a$ Wyckoff position (0, 0, 0) to the $1e$ Wyckoff position (2/3, 1/3, 0) (see Fig. \ref{bands}). At ambient conditions, the formation energy of the $P\bar{6}m2(1e)$ structure is 48.2 meV ($\sim$500K) per atom higher than of the $P\bar{6}m2$ ground state structure, the same energy scale as the structural transition around $\sim$425 K observed in the high-$T$ XRD and TEM experiments\cite{Canfield}. Our calculations suggest that the resistivity hysteresis observed at low pressures is indeed associated with this subtle structural transition, in line with those reported by U. Kaluarachchi, \emph{et al}~\cite{Canfield}. We also note that the critical pressure from the calculations ($\sim$5 GPa) is overestimated by a factor of 20 compared with the experimentally observed value ($\sim$0.25 GPa). This overestimation is an inherent issue in DFT calculations. At higher pressures, our study further predicted a new structural transition from $P\bar{6}m2(1e)$ to the tetragonal structure with space group $P4/mmm$ above 30 GPa, revealing a second structural phase transition under pressure.

It is intriguing to see how the electronic structure changes with these structural transitions. The calculated band structures without and with SOC for different structures are shown in Fig.\ref{bands}. Our results for the non-centrosymmetric hexagonal $P\bar{6}m2$ (Fig.\ref{bands}(d),(g)) are in good agreement with previous reports\cite{Cava,Sankar,sciadv,ncomms10556}. Large SOC splitting is clearly visible at the $K$ and $H$ points in the Brillouin zone for both Pb and Ta orbitals. The electron-like Pb $p$ orbital and hole-like Ta $d$ orbital cross each other at the Fermi level, resulting in the nodal-line states at the $K$ and $H$ points, topologically protected by mirror reflection symmetry. Fig.\ref{bands}(e),(h) represent the band structure of $P\bar{6}m2(1e)$ phase. The band structure is overall similar to that in the $P\bar{6}m2$ except that the nodal line states around the $H$ point are completely destroyed. In this phase, nodal-line states at $K$ point are partially gapped out by SOC. Fig.\ref{bands}(f),(i) delineate the band structure of $P4/mmm$ phase which is completely distinct from the above two phases. A band crossing exists at $M$ point but with no band inversion (see (f)) and a gap is induced by SOC (see (i)), which suggests a topologically trivial phase in this high-pressure structure.

\begin{figure}
\includegraphics[width=8.2cm]{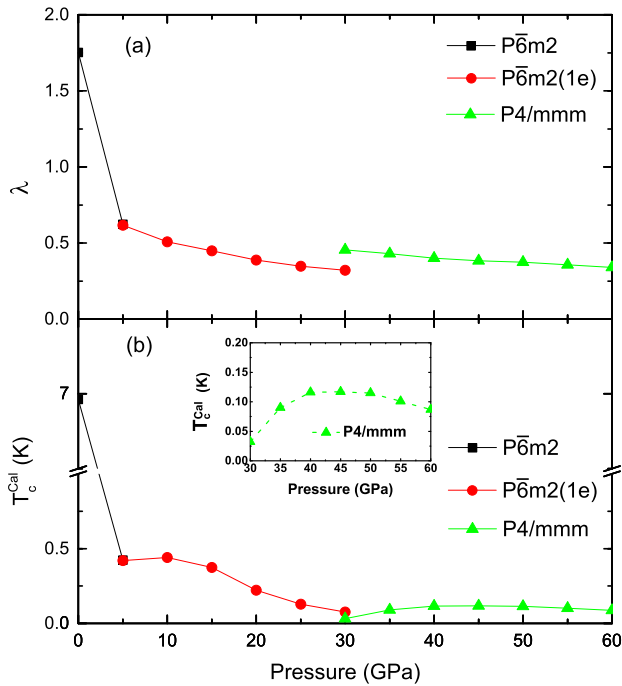}
\caption{\label{Tc} (a) Electron-phonon coupling constant $\lambda$, and (b) superconducting $T^{Cal}_{c}$
as a function of pressure. Inset shows the blow-up of the superconducting dome in $P4/mmm$ phase.}
\end{figure}

In order to estimate the superconducting characteristics of PbTaSe$_{2}$ under pressure, we performed linear response calculations~\cite{linear} of its electron-phonon properties, and estimated the critical temperature through the Mc-Millan Allen-Dynes formula~\cite{Mcmillan1968,Allen1975}

\begin{equation}
T^{Cal}_{c}=\frac{\omega_{ln}}{1.2}\textrm{exp}\left[-\frac{1.04(1+\lambda)}{\lambda-\mu^*(1+0.62\lambda)}\right],
\end{equation}

\noindent where $\omega_{ln}$ is the logarithmically averaged phonon frequency, and $\mu^*$ is the Coulomb pseudopotential which is set to be 0.1 in the calculations. We evaluate the pressure-dependent electron-phonon coupling constants $\lambda$ and the superconducting transition temperature $T^{Cal}_c$ shown in Fig.\ref{Tc}. It is found that both $\lambda$ and $T^{Cal}_c$ are maximal in $P\bar{6}m2$ phase at 0 GPa. Note that $T^{Cal}_c$ is twice the actual value of $T_c$ observed in the experiments. As the pressure increases, both $\lambda$ and $T^{Cal}_c$ decrease up to 30 GPa. For the $P4/mmm$ phase, while the electron-phonon coupling constant slightly decreases with pressure, $T^{Cal}_c$ demonstrates a superconducting dome, reaching a maximum value of 0.12 K at $\sim$45 GPa. However, this small value of $T_c$ makes the experimental verification very difficult.

\section{Conclusions}

In conclusion, we have studied the high-pressure superconducting phase diagram of PbTaSe$_{2}$ from both experiments and first-principles calculations. Superconductivity shows a marked suppression above $\sim$0.25 GPa, along with pronounced changes in the magnetoresistance and Hall resistivity, suggesting a Lifshitz transition under this pressure. The first-principles calculations unveil the structural origin for this Lifshitz transition, due to the shifting of the Pb atom from one Wyckoff coordinate to the other without changing its global symmetry. As a result, the nodal-line structure at the $K$ point is partially gapped while the Dirac states around the $H$ point are totally stripped away. The calculations further reveal a second structural transition at $\sim$30 GPa which is accompanied with a topological phase transition, from a topologically nontrivial state to a topologically trivial state. In this new state, a superconducting dome of electron-phonon interaction in its pairing origin is uncovered in the calculations and motivates the need for experimental investigations in due course.

\begin{acknowledgments}
The authors would like to thank Nigel Hussey, C. M. J. Andrew for the fruitful discussion. This work is sponsored by the National Key Basic Research Program of China (Grant No. 2014CB648400), and by National Natural Science Foundation of China (Grant No. 11474080, 11504182, 11374043), the Natural Science Foundation of Jiangsu Province (Grant No. BK20150831), Natural Science Foundation of Jiangsu Educational Department (Grant No. 15KJA430001), and six-talent peak of Jiangsu Province (Grants No. 2012-XCL-036) X.X. would also like to acknowledge the financial support from an open program from Wuhan National High Magnetic Field Center (2015KF15).

C. Q. Xu, R. Sankar and W. Zhou contributed equally to this work.

\end{acknowledgments}


\end{document}